\documentclass[aps,prl,reprint,groupedaddress,showpacs,floatfix]{revtex4-1}

\usepackage{graphicx}
\usepackage{dcolumn}
\usepackage{bm}
\usepackage[mathlines]{lineno}

\begin{document}

\title{\hfill {\small Phys. Rev. Lett. {\bf } (2014)}\\
       Semiconducting layered blue phosphorus:
       A computational study}

\author{Zhen Zhu}
\affiliation{Physics and Astronomy Department,
             Michigan State University,
             East Lansing, Michigan 48824, USA}

\author{David Tom\'{a}nek}
\email%
{tomanek@pa.msu.edu}%
\affiliation{Physics and Astronomy Department,
             Michigan State University,
             East Lansing, Michigan 48824, USA}

\date{\today} 

\begin{abstract}
We investigate a previously unknown phase of phosphorus that
shares its layered structure and high stability with the black
phosphorus allotrope. We find the in-plane hexagonal structure and
bulk layer stacking of this structure, which we call `blue
phosphorus', to be related to graphite. Unlike graphite and black
phosphorus, blue phosphorus displays a wide fundamental band gap.
Still, it should exfoliate easily to form quasi-2D structures
suitable for electronic applications. We study a likely
transformation pathway from black to blue phosphorus and discuss
possible ways to synthesize the new structure.
\end{abstract}




\maketitle

\begin{figure}[tb]
\includegraphics[width=1.0\columnwidth]{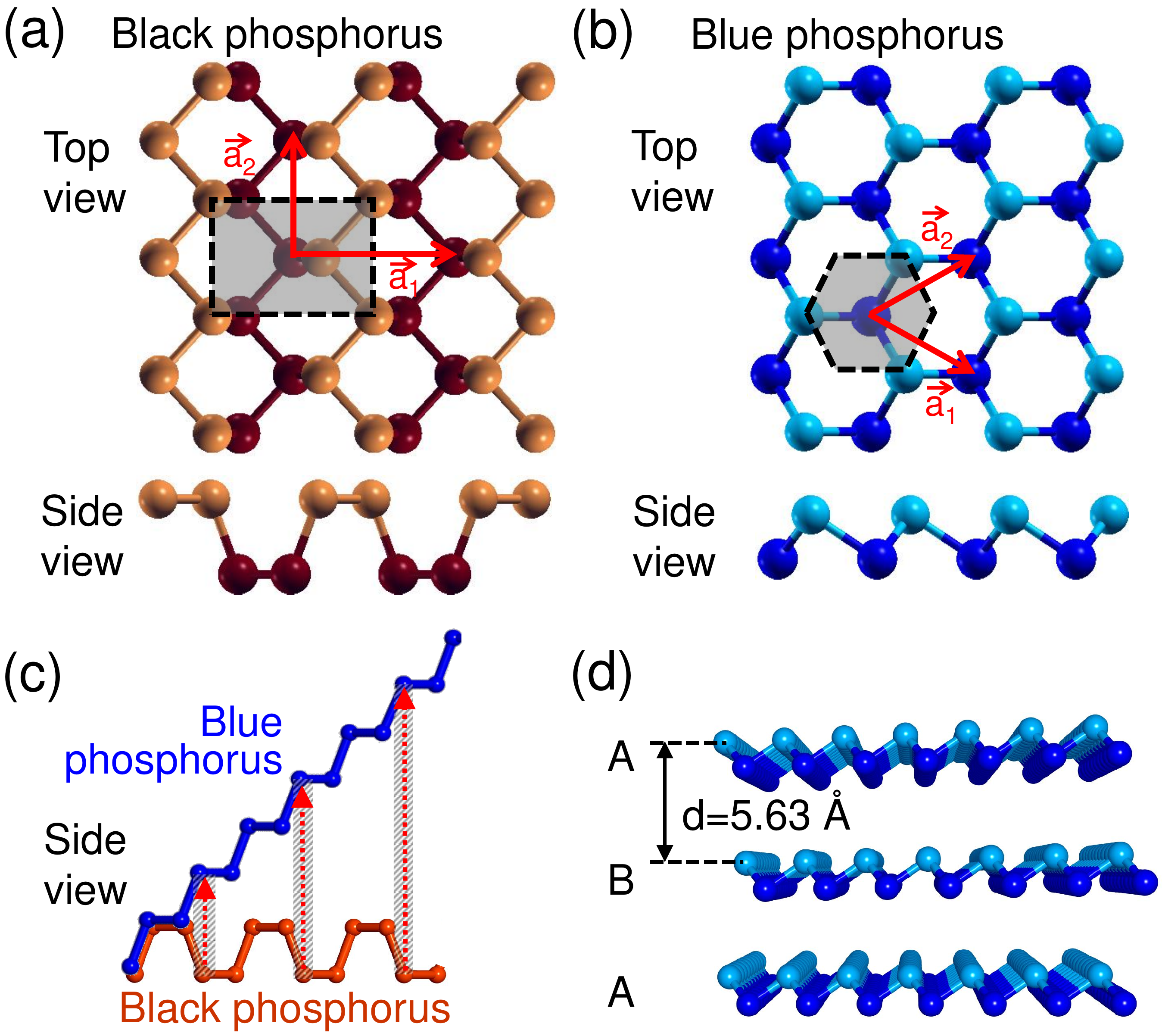}
\caption{(Color online) The layered structure of (a) black and (b)
blue phosphorus in top and side view. Atoms at the top and bottom
of the non-planar layers are distinguished by color and shading
and the Wigner-Seitz cells are shown by the shaded region. (c)
Schematic of the conversion of black to blue phosphorus by
dislocations, highlighted by the shaded regions and arrows. (d)
Equilibrium structure of AB stacked blue phosphorus in side view.
\label{fig1}}
\end{figure}

Elemental phosphorus is stable in a large number of structures,
including the common white, red, violet and black
allotropes\cite{{redp-blackp-phase1},{violetp}}, with the color
defined by the fundamental band gap. Most stable among them is
black phosphorus, which -- besides graphitic carbon -- is the only
layered structure of an elemental solid we know of. Individual
layers of black phosphorus, shown in Fig.~\ref{fig1}(a), resemble
the honeycomb structure of graphene in terms of connectivity, but
are non-planar. We noted that specific dislocations may convert
black phosphorus, characterized by armchair ridges in the side
view of the layers, to a well-defined structure with zigzag
puckering, shown in Fig.~\ref{fig1}(b). Assuming that the modified
structure is stable, it appears worthwhile to study the
equilibrium atomic arrangement in the bulk and the possibility of
exfoliating individual layers. Whereas the observed fundamental
band gap of 0.3-0.4~eV in black
phosphorus\cite{{Maruyama1981},{Narita1983}} is rather narrow, it
is intriguing to see if the modified phosphorus structure is a
semiconductor with a wider gap. If also the carrier mobility were
high, few-layer phosphorus in the new phase would become a worthy
contender in the emerging field of post-graphene 2D electronics.

Here we use {\em ab initio} calculations to investigate this
previously unknown phase of phosphorus that shares its layered
structure with the most stable black phosphorus allotrope. We find
this structural phase, which we call `blue phosphorus', to be
nearly as stable as black phosphorus. Whereas the in-plane
hexagonal structure and bulk layer stacking of blue phosphorus are
closely related to graphite, the main advantage of blue phosphorus
is its wide fundamental band gap in excess of 2~eV. Due to the
weak inter-layer interaction, blue phosphorus should exfoliate
easily to form quasi-2D structures for potential electronic
applications. We study a likely transformation pathway from black
to blue phosphorus and discuss possible ways to synthesize the
postulated structure.


Our computational approach to gain insight into the equilibrium
structure, stability and electronic properties of blue phosphorus
is based on {\em ab initio} density functional theory (DFT) as
implemented in the \textsc{SIESTA}~\cite{SIESTA} and
VASP\cite{VASP} codes. We used periodic boundary conditions
throughout the study, with multilayer structures represented by a
periodic array of slabs separated by a 15~{\AA} thick vacuum
region. We used the Perdew-Burke-Ernzerhof~\cite{PBE}
exchange-correlation functional, norm-conserving Troullier-Martins
pseudopotentials~\cite{Troullier91}, and a double-$\zeta$ basis
including polarization orbitals. The reciprocal space was sampled
by a fine grid~\cite{Monkhorst-Pack76} of
$8{\times}8{\times}1$~$k$-points in the Brillouin zone of the
primitive unit cell. We used a mesh cutoff energy of $180$~Ry to
determine the self-consistent charge density, which provided us
with a precision in total energy of ${\alt}2$~meV/atom. All
geometries have been optimized by \textsc{SIESTA} using the
conjugate gradient method\cite{CGmethod}, until none of the
residual Hellmann-Feynman forces exceeded $10^{-2}$~eV/{\AA}. Our
\textsc{SIESTA} results for the optimized geometry, interlayer
interactions and electronic structure were found to be in general
agreement with \textsc{VASP} calculations. To verify the stability
of the system at elevated temperatures, we performed a canonical
molecular dynamics calculation of the monolayer and free-standing
flakes using the \textsc{SIESTA} code. We used the Verlet
integration algorithm to cover time periods up to 1~ps with the
time step of 1~fs and present our results in the Supplemental
Material\cite{SM-pblue14}.


The optimized reference structure of black phosphorus, shown in
Fig.~\ref{fig1}(a), is presented next to the proposed blue
phosphorus structure in Fig.~\ref{fig1}(b). The top view of both
structures illustrates their similarity with the honeycomb lattice
of graphite, which contains two atoms per layer per unit cell.
Both phosphorus allotropes differ from graphite in the non-planar
structure of their layers. In top view, the isotropic structure of
blue phosphorus in Fig.~\ref{fig1}(b) differs significantly from
the anisotropic structure of black phosphorus in
Fig.~\ref{fig1}(a). As seen in side view in Figs.~\ref{fig1}(a)
and \ref{fig1}(b), the puckered zigzag structure in the
cross-section of blue phosphorus differs from the distinct
armchair ridges that cause the anisotropy of black phosphorus. The
puckering in the blue phosphorus monolayer is similar to the
postulated structure of single-wall phosphorus
nanotubes\cite{Seifert2000}.

The structural relationship between blue and black phosphorus is
illustrated schematically in Fig.~\ref{fig1}(c). A dislocation is
introduced in a monolayer of black phosphorus by flipping specific
P atoms from a `down' to an `up' position without changing the
local bond angles, as described below and indicated by the arrows
in Fig.~\ref{fig1}(c). Subjecting every fourth row of P atoms to
this transformation converts a monolayer of black to blue
phosphorus. The location of dislocation lines in the monolayer
structures is emphasized by the shaded regions in
Fig.~\ref{fig1}(c).

Atoms in the layers of blue phosphorus are covalently bonded at
the equilibrium distance of 2.27~{\AA}, resulting in a large
binding energy of 5.19~eV/atom. This value differs from that of
black phosphorus by less than $2$~meV/atom, suggesting that blue
and black phosphorus are equally stable.
A weak inter-layer interaction\cite{SM-pblue14} of 6~meV/atom
holds the layered structure together at the interlayer distance
$d=5.63$~{\AA}, as seen in Fig.~\ref{fig1}(d), which indicates the
possibility of easy exfoliation. The AB hexagonal stacking and the
ABC rhombohedral stacking of layers differ energetically by less
than $1$~meV/atom.

The optimized hexagonal unit cell of an isolated blue phosphorus
monolayer, shown in Fig.~\ref{fig1}(b), is spanned by lattice
vectors $\vec{a}_1$ and $\vec{a}_2$, with
$a=|\vec{a}_1|=|\vec{a}_2|=3.33$~{\AA}. The influence of the
interlayer interaction on the in-layer structure is small, causing
only a negligible change from $a=3.324$~{\AA} in the bulk to
$a=3.326$~{\AA} in the isolated monolayer. With its higher
symmetry, the smaller hexagonal Wigner-Seitz cell of blue
phosphorus, which contains two atoms and is shown in
Fig.~\ref{fig1}(b), differs from the rectangular Wigner-Seitz cell
of the anisotropic black phosphorus monolayer with 4 atoms/unit
cell, as shown in Fig.~\ref{fig1}(a). Still, monolayers of blue
and black phosphorus may form an `ideal' in-layer connection by
the dislocation illustrated in Fig.~\ref{fig1}(c).

\begin{figure}[tb]
\includegraphics[width=1.0\columnwidth]{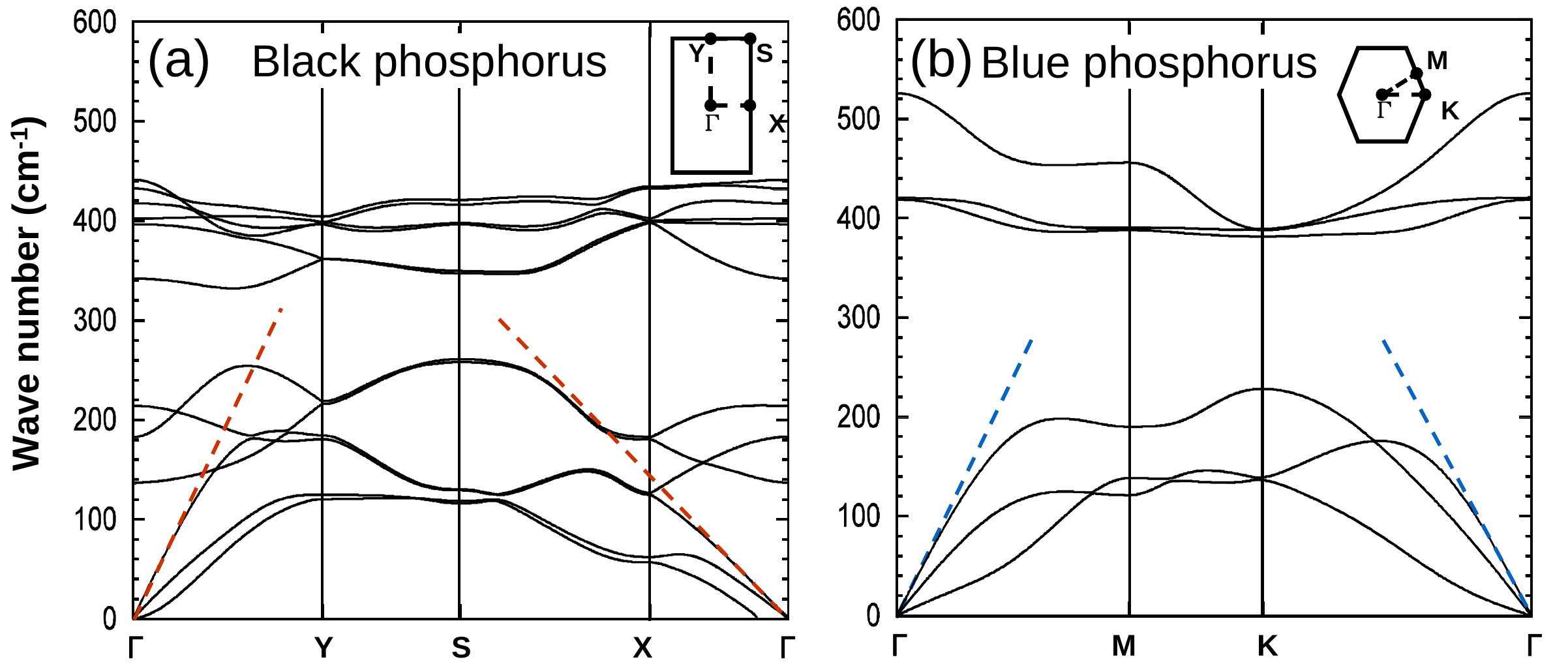}
\caption{(Color online) Vibrational band structure
$\omega(\vec{k})$ of a monolayer of (a) black and (b) blue
phosphorus. The slope of the dashed lines along the longitudinal
acoustic branches near $\Gamma$ corresponds to the speed of sound
and the in-plane stiffness. \label{fig2}}
\end{figure}


One way to compare the stability and structural rigidity of the
different phosphorus allotropes is by studying the vibration
spectrum. Our results for the vibration spectra of blue and black
phosphorus monolayers are presented in Fig.~\ref{fig2}. We find
the vibration spectra to be rather similar, reflecting a very
similar bonding character. Acoustic and optical modes are well
separated in both black and blue phosphorus. The harder
longitudinal optical modes reflect a higher in-plane rigidity of
the blue phosphorus allotrope in comparison to the accordion-like
black phosphorus structure. The calculated flexural rigidity value
$D=0.84$~eV in blue phosphorus is lower than the $D=1.51$~eV value
in black phosphorus. These values provide a quantitative
explanation for the dispersion of the flexural acoustic modes ZA
near the $\Gamma$ point\cite{Sanchez-Portal-CNT1999} in the two
allotropes\cite{SM-pblue14}. The high rigidity of a free-standing
blue phosphorus monolayer was also confirmed by our molecular
dynamics calculations at nonzero temperatures, discussed in the
Supplemental Material\cite{SM-pblue14}.

\begin{figure*}[t]
\includegraphics[width=1.4\columnwidth]{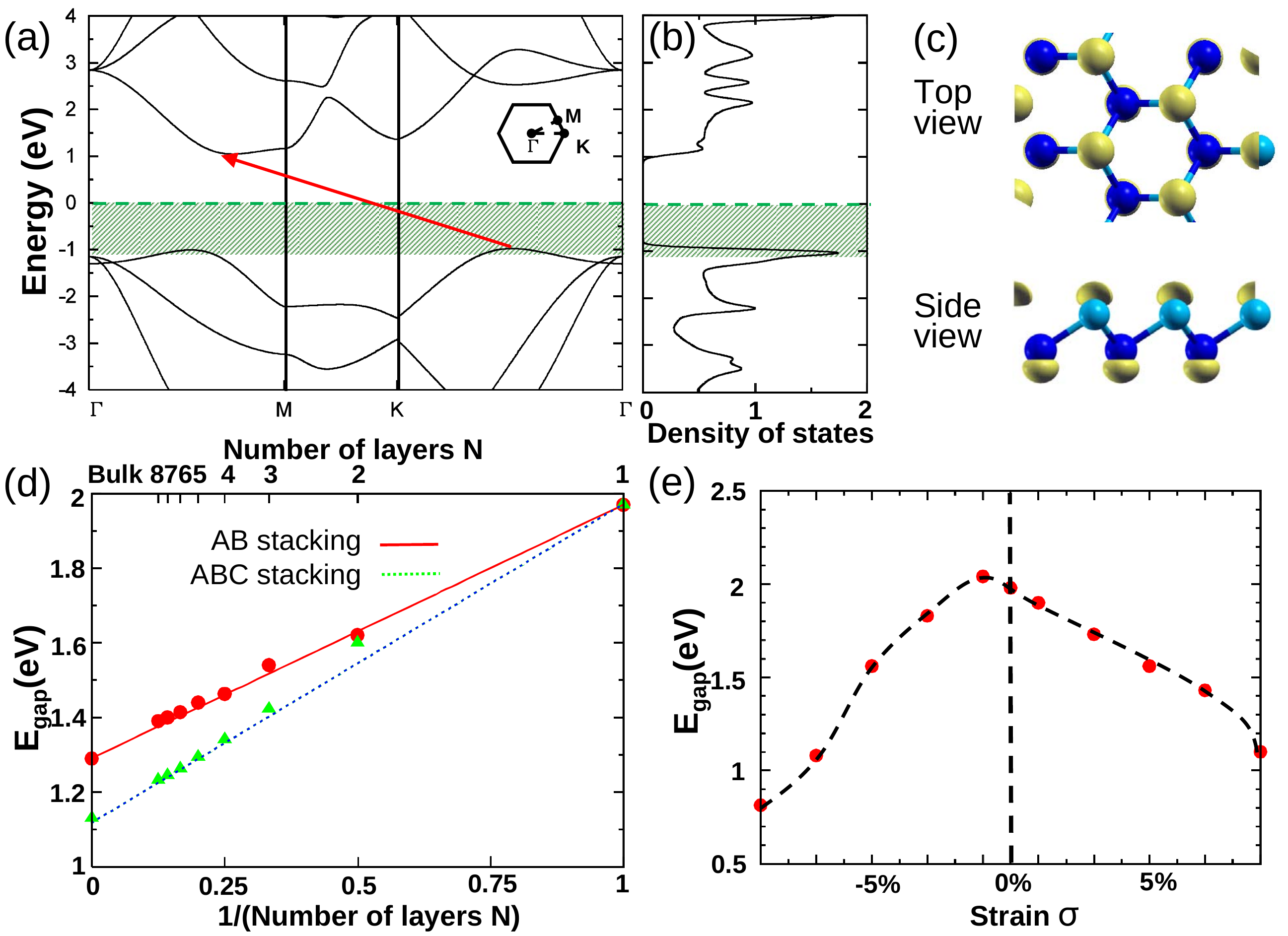}
\caption{(Color online) (a) Electronic band structure and (b)
density of states of a blue phosphorus monolayer. (c) Electron
density $\rho_{vb}$ in the 0.1~eV wide energy range near the top
top of the valence band in blue phosphorus, indicated by the green
shaded region in (a) and (b). $\rho_{vb}$ is represented at the
isosurface value $\rho=7{\times}10^{-3}$~e/{\AA}$^3$ and
superposed with a ball-and-stick model of the structure.
Dependence of the fundamental band gap $E_g$ on (d) the number of
layers and (e) the in-plane stretching along the $\vec{a}_1$
direction. \label{fig3}}
\end{figure*}

We next compare the slopes of the longitudinal acoustic branches
near $\Gamma$, which correspond to the speed of sound and reveal
the in-plane stiffness. As seen in Fig.~\ref{fig2}(a), the speed
of sound along the $\Gamma-Y$ direction in black phosphorus,
$v_s^{\Gamma-Y}=7.8$~km/s, is significantly higher than the
$v_s^{\Gamma-X}=3.8$~km/s value along the $\Gamma-X$ direction,
reflecting an anisotropy in the elastic constants. The lower
rigidity along the $\Gamma-X$ direction, corresponding to the
$\vec{a}_1$ direction in Fig.~\ref{fig1}(a), reflects the fact
that compression along $\vec{a}_1$ requires primarily bond
bending, which comes at a lower energy cost than bond stretching.
In strong contrast to those findings, our results in
Fig.~\ref{fig2}(b) indicate that the in-plane elastic response of
blue phosphorus is nearly isotropic, with nearly the same value
$v_s=7.7$~km/s for the speed of sound along the $\Gamma-M$ and the
$\Gamma-K$ direction. The predicted finite in-layer
compressibility of these non-planar structures is advantageous
when accommodating lattice mismatch during Chemical Vapor
Deposition (CVD) growth on a substrate.


Our DFT results for the electronic structure of blue phosphorus
are presented in Fig.~\ref{fig3}. The calculated band structure
and density of states, presented in Figs.~\ref{fig3}(a) and
\ref{fig3}(b), indicate that a blue phosphorus monolayer is a
semiconductor with an indirect band gap $E_{gap}{\approx}2$~eV.
Comparison with more proper self-energy calculations based on the
GW approach, performed by \textsc{VASP}, indicate that the DFT
band gap is underestimated by ${\approx}1.0{\pm}0.2$~eV in mono-
and multi-layers of blue P as a common shortcoming of DFT. Still,
the electronic structure of the valence and the conduction band
region in DFT is believed to closely represent experimental
results. Therefore, we expect the charge density associated with
states near the top of the valence band, shown in
Fig.~\ref{fig3}(c), to be represented accurately. These states
correspond to the energy range highlighted by the green shading in
Figs.~\ref{fig3}(a) and \ref{fig3}(b), which extends from mid-gap
to 0.1~eV below the top of the valence band. These states cause
the inter-layer band dispersion in few-layer
systems\cite{SM-pblue14}, and their hybrids with electronic states
of the contact electrodes will play a crucial role in the carrier
injection and quantum transport.

More interesting than the precise value of the fundamental band
gap is its dependence on the number of layers $N$ in a multi-layer
slab, shown in Fig.~\ref{fig3}(d). This is a consequence of the
inter-layer dispersion near the Fermi level in the bulk material,
which is discussed in the Supplemental Material\cite{SM-pblue14}.
Independent of the type of stacking, we find $E_{gap}$ to be
inversely proportional to $N$ between one monolayer and the bulk
structure. We conclude that modifying the slab thickness may
change the value of $E_{gap}$ by up to a factor of 2, which is
very important for electronic applications. The $N$-dependent band
gap value of ${\alt}3$~eV
lies just above the photon energy of visible blue light. We derive
the name `blue phosphorus' from this absorption edge, which plays
a key role in the optical appearance.

As seen in Fig.~\ref{fig3}(e), the band gap $E_{gap}$ depends also
sensitively on the applied in-layer strain $\sigma$. In view of
the non-planarity of the structure, the discussed strain range
$-10\%<\sigma<+10\%$ can be achieved without a large energy
penalty, as discussed in connection with the vibration spectra.
The possibility to change the band gap value by ${\alt}50\%$ in
strained epitaxial geometries on different substrates is one more
indication that blue phosphorus may find intriguing applications
in nanoelectronics.

\begin{figure}[tb]
\includegraphics[width=1.0\columnwidth]{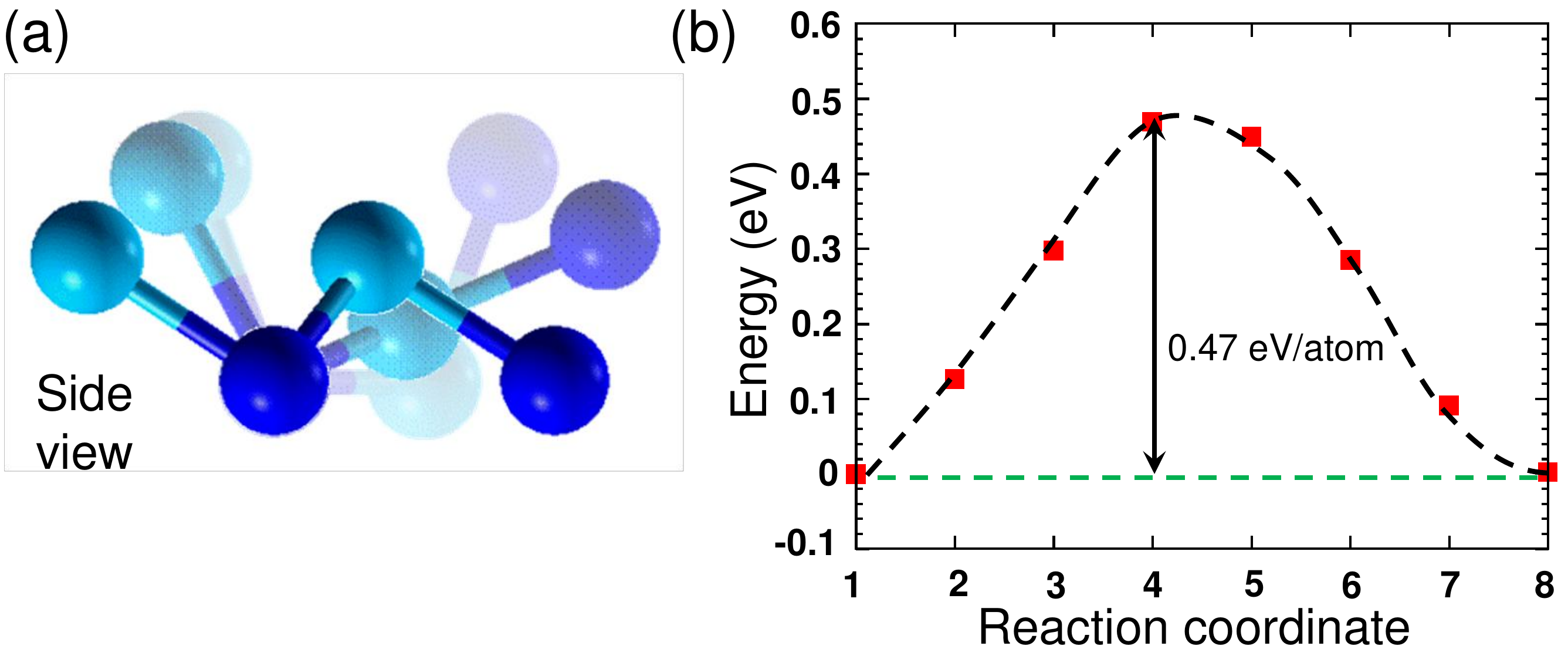}
\caption{(Color online) (a) Series of structural snapshots
depicting the transition from black to blue phosphorus. (b) Total
energy change during the transformation from black to blue
phosphorus. The reaction coordinate range `1-2' describes initial
in-layer stretching along the horizontal direction, identified as
$\vec{a}_1$ in Fig.~\ref{fig1}(a), followed by out-of-plane
displacement of atoms at a fixed value of $\vec{a}_1$ in the range
`2-8'. \label{fig4}}
\end{figure}


We next studied the possibility of converting a monolayer of black
to blue phosphorus by introducing an array of dislocations,
depicted schematically in Fig.~\ref{fig1}(c). Our discussion of
the conversion process including energy estimates is presented in
Fig.~\ref{fig4} and the Supplemental Material\cite{SM-pblue14}.
The starting black phosphorus, the final blue phosphorus, and an
intermediate structure are depicted in side view in
Fig.~\ref{fig4}(a). To estimate changes in the atomic arrangement
during the conversion process, we changed and constrained the
out-of-plane displacement of specific atoms in the unit cell to
follow the `black-to-blue' trajectory and relaxed all other
degrees of freedom. This provided us with a sequence of structures
`1-8' in-between black and blue phosphorus, which loosely define
the reaction coordinate in Fig.~\ref{fig4}(b). The change in the
unit cell from black to blue phosphorus has been imposed between
steps `1' and `2' by deforming it from the initial size and shape
in black phosphorus to the final rectangular supercell in blue
phosphorus, followed by atomic relaxation. Our results for the
relative energy with respect to the black phosphorus structure `1'
again illustrate our finding that the blue phosphorus structure
`8' is equally stable as `1'. The activation barrier for the
conversion process is likely overestimated due to the constraints
imposed on the intermediate structures and may further be lowered
by stretching black phosphorus layers. Thus the true value should
be below the already low value of $0.47$~eV/atom, indicating the
relative ease of mechanical conversion from black to blue
phosphorus.

As suggested earlier, the weak inter-layer interaction should
allow mechanical exfoliation of blue phosphorus in analogy to the
black allotrope\cite{PYe2014}. As a matter of fact, depositing
mechanically a monolayer of black phosphorus onto a stepped
substrate may cause formation of dislocation lines and thus
formation of narrow domains of blue phosphorus in the monolayer
structure along the step edges of the substrate. In analogy to
graphene\cite{{KimNat2009},{ReinaNL2009}} and
silicene\cite{VogtPRL2012}, also layered phosphorus structures may
be grown by CVD on specific substrates. As suggested in this
study, blue and black phosphorus should be equally stable. The
preferential phase should thus be determined by the lattice
constant and the symmetry of the substrate in order to maximize
the adsorption energy. Consequently, we expect blue phosphorus to
form preferentially on substrates with hexagonal symmetry and a
matching lattice constant, such as MoS$_2$, or the (0001) surfaces
of Zr and Sc, whereas black phosphorus should start growing on
substrates with a rectangular lattice. Due to the low energy of
forming a dislocation that connects blue and black phosphorus,
both allotropes could coexist on particular substrates, including
stepped surfaces, to optimize the adlayer-substrate interaction.

One of the main reasons for the interest in 2D semiconductors
including graphene for electronic applications is the observed
high mobility of carriers. Related quasi-2D systems, including
MoS$_2$, bring the benefit of a nonzero band gap, but display
lower intrinsic carrier mobility due to enhanced electron-phonon
coupling, primarily caused by the presence of heavy elements such
as Mo\cite{DT221}. It appears likely that the blue phosphorus
structure, similar to black phosphorus\cite{PYe2014}, may exhibit
a higher carrier mobility than MoS$_2$. The combination of a
significant band gap and high carrier mobility would turn blue
phosphorus into an excellent contender for a new generation of
nano-electronic devices.


In conclusion, we have conducted {\em ab initio} calculations to
investigate a previously unknown layered phase of phosphorus,
which we call `blue phosphorus'. We find blue phosphorus to be
nearly as stable as black phosphorus, the most stable phosphorus
allotrope. While sharing the atomic connectivity with the
honeycomb lattice of graphene, layers of blue phosphorus are
non-planar. Whereas the bulk layer stacking of blue phosphorus is
closely related to graphite, the main advantage of this allotrope
is its wide fundamental band gap in excess of 2~eV. Due to the
weak inter-layer interaction, blue phosphorus should exfoliate
easily to form quasi-2D structures for potential electronic
applications. We have investigated a likely transformation pathway
from black to blue phosphorus and show that the postulated
structure may form spontaneously by CVD on a lattice-matched
substrate or may be result by stretching black phosphorus.
Monolayers of blue phosphorus may form a structurally ideal
connection to monolayers of black phosphorus with a different
electronic structure.

\begin{acknowledgements}
We thank Peide Ye for directing our attention towards the related
black phosphorus structure. This study was supported by the
National Science Foundation Cooperative Agreement \#EEC-0832785,
titled ``NSEC: Center for High-rate Nanomanufacturing''.
Computational resources have been provided by the Michigan State
University High Performance Computing Center.
\end{acknowledgements}

{\it Note added in proof.}---We have learned that blue phosphorus
is structurally related to the postulated\cite{Boulfelfel12} and
observed\cite{Jamieson63} $A7$ phase of phosphorus.



%

\end{document}